# Modern Portfolio Theory using SAS® OR

Murphy Choy, School of Information System, SMU, Singapore

## ABSTRACT

Investment approaches in financial instruments have been varied and often produce unpredictable results. Many investors in the earlier days of investment banking suffered catastrophical losses due to poor strategy and lack of understanding of the financial market. With the development of investment banking, many innovative investment strategies have been proposed to make portfolio returns higher than the overall market. One of the most famous theories of portfolio creation and management is the modern portfolio theory proposed by Harry Markowitz. In this paper, we shall apply the theory in creating a portfolio of stocks as well as managing it.

## INTRODUCTION

Investment banking has been one of the most interesting branches of banking for a variety of reasons ranging from the extreme levels of salary and bonus one receive to the complicated strategies in executing the various investment strategies in different financial markets. As many of the financial trades are done in environment where common people are unable to access, many of the more sophisticated financial instruments are not available for common people to purchase as part of their portfolio. For example, there are not many people on the street who will purchase a futures contract of Euro worth $125,000. However, the common people will be likely to acquire a small portfolio consisting of stocks and warrants. There are many approaches to creating a profitable portfolio. Given the experience and knowledge of the individual, one can choose from the simple dollar time averaging strategies to the complex day trading techniques. For the SAS programmers, our access to SAS allows us a simple way to manage our little nest of golden eggs with some programming and modern portfolio theory.

## MODERN PORTFOLIO THEORY

The modern portfolio theory began with a paper (Markowitz, 1952) and book (Markowitz, 1959) written by Harry Markowitz. The fundamental premise of modern portfolio theory is that any stocks has a probability to go up or down depending on the market and thus their inclusion or exclusion in a portfolio does not matter individually. However, when they are placed together, the interaction between the stocks reduces the overall price volatility which then contributes to the stability of the portfolio. Any portfolio is designed with returns in mind and with the modern portfolio theory, one can choose an expected return and then seek to minimize the risk (volatility) associated with the combination of stocks.

## ASSUMPTIONS

There are several assumptions about the modern portfolio theory which are listed below (less than the original one).

1. Returns from the portfolio is normally distributed (multivariate normality is assumed).
2. Correlations between the stocks are fixed or constant for a period of time.
3. The investors seek to maximize their overall profit/economic utility.
4. All players in the market are rational and risk adverse.
5. Common information is available to all players in the market.
6. All players are price takers.

While many of the assumptions are objectionable, we will discuss the short comings of the modern portfolio theory in later sections and how we can overcome it in theory and SAS.

## MATHEMATICAL FORMULATION

Assume the following annotation,

$w_i$ = weight allocation to each individual stock *i* in the portfolio.

$a_i$ = return of each individual stock *i* in the portfolio.

$\sigma_i$ = volatility of each individual stock *i* in the portfolio.

$\rho_{ij}$ = correlation coefficient between stock *i* and stock *j* in the portfolio.

Modern Portfolio Theory using SAS OR, continued

Expected Returns

$$E(a_p) = \sum_{i=1}^{n} w_i a_i$$

Volatility of portfolio

$$\sigma_p = \sqrt{\sum_{i=1}^{n} w_i^2 \sigma_i^2 + \sum_{i=1}^{n} \sum_{i \neq j}^{n} w_i w_j \sigma_i \sigma_j \rho_{ij}}$$

## OPTIMIZATION OF THE PORTFOLIO

One of the key problems faced by the modern portfolio theory is the selection of the stocks in a way to reduce the volatility while maintain an acceptable level of returns. This problem is an optimization problem and can be solved using quadratic programming. Quadratic programming is an advanced form of linear programming where the linearity assumption has been relaxed as the problem formulation requires a quadratic calculation. Below is the mathematical formulation of the problem.

Objective: Minimize $\sigma_p$

Subjected to the following constraints:

Constraint 1 (Returns constraint): $E(a_p) \geq P(r)$

Constraint 2 (Budget constraint): $\sum_{i=1}^{n} w_i a_i \leq Budget$

The constraints are fairly obvious to the portfolio creator. For any investors, they have a limited budget in which they are willing to invest given a sufficient level of return that they will receive. By reducing the volatility, they can expect that the return will not fluctuate too wildly leading to unpredictable level of return. However, occasionally, some extremely conservative investors wish to further limit their possible loss and add in a limiter constraint.

Objective: Minimize $\sigma_p$

Subjected to the following constraints:

Constraint 1 (Returns constraint): $E(a_p) \geq P(r)$

Constraint 2 (Budget constraint): $\sum_{i=1}^{n} w_i a_i \leq Budget$

Constraint 3 (Limit constraint): $E(a_p) - (3 \times \sigma_p) \geq 0$

In certain environment, certain stock positions are not possible and thus that has to be factored in as well. This is particularly the case during the financial crisis of 2008-2009 where shorting of stocks is barred in certain markets.

Objective: Minimize $\sigma_p$

Subjected to the following constraints:

Constraint 1 (Returns constraint): $E(a_p) \geq P(r)$

Constraint 2 (Budget constraint): $\sum_{i=1}^{n} w_i a_i \leq Budget$

Constraint 3 (Position constraint): $w_i \geq 0$

The optimization of the portfolio can be subjected to many different constraints that are dependent on the environment that it operates in. The flexibility of the approach makes it extremely attractive to the any investors with good access to excel spreadsheets. However, it is precisely this simplicity that causes trouble for the users of the technique. Most users who implement the technique in excel spreadsheet tends to over restrict the number of stocks available for analysis. This leads to a severe shortage of possible combination which can reduce the overall volatility or contribute to the return. At the same time, the information that can be contained in an excel spreadsheet will be limited by the computational abilities of excel. There are several programming languages adopted in portfolio management and creation. We will cover the approach using SAS.

## GETTING THE RIGHT INFORMATION

Information is the most critical section of the modeling. The classic phrase 'garbage in, garbage out' is the most relevant here. SAS has several data importing abilities that allow it to process information from the internet in an efficient manner. This is one very important use of SAS in data gathering especially when one needs to build a portfolio of stocks using information collated from many places. In this paper, I will demonstrate the use of recursive code calling to extract data.

One of the easiest places to obtain massive amount of stock information is from Yahoo! Finance. Yahoo! Finance provides a free service where the stock ticks are compiled in the form of CSV text files which can be read into SAS easily. However, manually obtaining information from Yahoo! Site will be disastrous given the huge number of stocks online. The fastest way to obtain all these information is to play around with the URL. Below is an example of a URL from Yahoo! for historical stock information for Apple.

http://ichart.finance.yahoo.com/table.csv?s=AAPL&d=3&e=16&f=2011&g=d&a=7&b=14&c=1986&ignore=.csv

The URL can be broken into its constituent portion. Below is the explanation.

http://ichart.finance.yahoo.com/table.csv?s=<StockTicketName>&d=<MonthToday>&e=<DayToday>&f=<YearToday>&g=d&a=7&b=14&c=1986&ignore=.csv

To basically extract the CSV, we only need to give the stock ticket name and the date information to reference the CSV files. However, as mentioned earlier, it is very important to be able to have the list of stock ticks to be able to call upon Yahoo! to provide the data. In this case, we are very fortunate to have a website which provides stock ticks for NASDAQ.

http://www.nasdaqtrader.com/dynamic/SymDir/nasdaqlisted.txt

This file contains all the stock ticks that can be found on NASDAQ. We can import all these information using the codes below.

```
FILENAME NASDQ URL 'http://www.nasdaqtrader.com/dynamic/SymDir/nasdaqlisted.txt';

DATA NASDAQSTOCKLIST;
INFILE NASDQ DELIMITER = '|' MISSOVER LRECL = 128 FIRSTOBS = 2;
LENGTH SYMBOL $5. SECURITYNAME $55. MARKETCATEGORY $15. TESTISSUE $2. FINANCIALSTATUS $2. ROUND $2.;
INPUT SYMBOL $ SECURITYNAME $ MARKETCATEGORY ~ TESTISSUE $ FINANCIALSTATUS $ ROUND $;
RUN;
```

Once the information has been collected, we can now attempt to automate the entire process of downloading and processing the historical stock information for portfolio creation. Automating the call can be easily done via a recursive call using the data set as the source of calling information. Using the combination of CALL EXECUTE and data, we can automate the process easily.

```
DATA _NULL_;
SET NASDAQSTOCKLIST;

CALL EXECUTE("filename "||Symbol||" url
'"||'http://ichart.finance.yahoo.com/table.csv?s='||compress(Symbol)||'&d=3&e=16&f=2011&g=d&a=7&b=14&c=1986&ignore=.csv'||'" DEBUG;");
CALL EXECUTE("data "||Symbol||";");
CALL EXECUTE("infile "||Symbol||" dsd lrecl = 128 firstobs = 2;");
CALL EXECUTE("informat Date yymmdd10.;");
CALL EXECUTE("input Date Open High Low Close Volume AdjClose;");
CALL EXECUTE("format Date yymmdd10.;");
CALL EXECUTE("RUN;");

RUN;
```

However, due to the amount of information being downloaded, this will take some time to finish even with a high bandwidth internet service. Once all the stock information has been downloaded, we can then process the data and prepare it for further analysis. For the sake of simplicity, we will be modeling using the closing price for each stock at the end of each day. With so many data sets in the library, one will seek a simple way to combine the files together. One simple trick is to use PROC CONTENTS with the _ALL_ keyword to list all the data tables in the library for management.

```sas
PROC SQL;

        DROP TABLE NASDAQSTOCKLIST;

QUIT;

PROC CONTENTS DATA = WORK._ALL_ OUT = CONTENTS NOPRINT;RUN;

PROC SORT DATA = CONTENTS(WHERE = (NOBS > 1000) KEEP = MEMNAME NOBS) NODUPKEY;
BY MEMNAME;
RUN;

DATA _NULL_;
SET CONTENTS;

CALL EXECUTE("PROC SORT DATA = "||MEMNAME||"(KEEP = DATE CLOSE);");
CALL EXECUTE("BY DATE;RUN;");

RUN;

DATA _NULL_;
SET CONTENTS END = EOF;

IF _N_ = 1 THEN CALL Execute("DATA OVERALL;MERGE");
CALL EXECUTE(MEMNAME||"(RENAME = (CLOSE = "||MEMNAME||"))");
IF EOF THEN DO;
        CALL EXECUTE(";BY DATE;RUN;");
END;
RUN;
```

Using the output from PROC CONTENTS, one can easily make use of the CALL EXECUTE ability to generate all the codes needed to run the process. The CALL EXECUTE combo allows for a massive amount of reduction of coding work needed. With the data prepared, we can now move on to the next step of modeling.

### CORRELATION MATRIX ESTIMATION

Estimation of the correlation matrix is one of the key steps to portfolio creation. The modern portfolio theory uses the correlation matrix to calculate the portfolio variance. The creation of correlation matrix is extremely simple in SAS. PROC CORR is the classic procedure to generate such a matrix. However, one could also use the covariance matrix to calculate the portfolio variance. PROC CORR provides the option to do either one and in this paper, for less computational work, I will be using the covariance directly.

```sas
PROC CORR DATA = <Stock data> OUT = CORRTABLE (WHERE=(UPCASE(_TYPE_) IN ("COV","MEAN"))) COV NOSIMPLE NOPRINT;
        VAR <Stocks>;
        WITH <Stocks>;
RUN;
```

At the same time, PROC CORR generate the means of the stocks as well which is needed as an input to the model. To use all the information available, we can break the table into its constituent component which can then be fed into the models.

```sas
DATA COVTABLE MEANTABLE;
SET CORRTABLE;
IF UPCASE(_TYPE_) IN ("MEAN") THEN OUTPUT MEANTABLE;
ELSE OUTPUT COVTABLE;
RUN;
```

Modern Portfolio Theory using SAS OR,continued

## OPTIMIZING THE PORTFOLIO

With the information in place, we can start looking at optimizing the portfolio. The optimization of the portfolio involves the use of quadratic optimization. As this is a basic introduction to portfolio creation, we will stick to the following assumptions and constraints.

Objective: Minimize $\sigma_p$

Subjected to the following constraints:

Constraint 1 (Returns constraint): $E(a_p) \geq P(r)$

Constraint 2 (Budget constraint): $\sum_{i=1}^{n} w_i a_i \leq Budget$

Basically in this scenario, we are assuming limited budget with an expected return in which we can have both long and short positions. While there are many different scenarios that might occur, this is subjected to the individual's preferences about his investment. PROC OPTMODEL allows one to easily manage the constraints as it incorporate the various types of optimization algorithm that are used given certain requirements are met. The syntax is as follow.

```
PROC OPTMODEL;

VAR  X{1..&VARCOUNT} >= 0;

NUM COEFF{1..&VARCOUNT, 1..&VARCOUNT} = [
        %DO I = 1 %TO &VARCOUNT;
                %DO J = 1 %TO &VARCOUNT;
                        &&COEFF&I&J
                %END;
        %END;
];

NUM R{1..&VARCOUNT}=[
                %DO I = 1 %TO &VARCOUNT;
                        &&MEAN&I
                %END;
];

/* MINIMIZE THE VARIANCE OF THE PORTFOLIO'S TOTAL RETURN */
MINIMIZE F = SUM{I IN 1..&VARCOUNT, J IN 1..&VARCOUNT}COEFF[I,J]*X[I]*X[J];

/* SUBJECT TO THE FOLLOWING CONSTRAINTS */
CON BUDGET: SUM{I IN 1..&VARCOUNT}X[I] = &BUDGET;
CON GROWTH: SUM{I IN 1..&VARCOUNT}R[I]*X[I] >= &RETURN_VALUE;

%IF &TYPE ^= S %THEN %DO;
        SOLVE WITH QP;

        PRINT X;
%END;
%ELSE %DO;
        FOR {I IN 1..&VARCOUNT} X[I].LB=-X[I].UB;

        SOLVE WITH QP;
%END;

PRINT X;

QUIT;
```

As one will notice that the syntax is done up with macro variables in many places, the main driver for this is the need of a macro that calculate the weights in the portfolio automatically by feeding it with the data of stocks information.

**PUTTING THESE TOGETHER**

A SAS Macro has been developed to achieve this. You can find it in the appendix A.

**CONCLUSION**

Portfolio creation is a very simple task in SAS given the abilities of SAS to access web information as well as excellent optimization routines.

**REFERENCE**

Markowitz, H.M. (March 1952). "Portfolio Selection". *The Journal of Finance* **7** (1): 77–91. doi:10.2307/2975974. JSTOR 2975974.

Markowitz, H.M. (1959). *Portfolio Selection: Efficient Diversification of Investments*. New York: John Wiley & Sons. (reprinted by Yale University Press, 1970, ISBN 978-0-300-01372-6; 2nd ed. Basil Blackwell, 1991, ISBN 978-1-55786-108-5)

**CONTACT INFORMATION**

Your comments and questions are valued and encouraged. Contact the author at:

>Name: Murphy Choy
>Enterprise: School of Information Systems, Singapore Management University
>Address: 80 Stamford Road
>City, State ZIP: Singapore 178902
>Work Phone: +65-92384058
>E-mail: goladin@gmail.com/murphychoy@smu.edu.sg

**ACKNOWLEDGEMENT**

SAS and all other SAS Institute Inc. product or service names are registered trademarks or trademarks of SAS Institute Inc. in the USA and other countries. ® indicates USA registration.  Other brand and product names are trademarks of their respective companies.

Modern Portfolio Theory using SAS OR,continued

**Appendix A**

```
/****************************************************************************
MODERN PORTFOLIO THEORY USING SAS/OR
****************************************************************************/

LIBNAME DATA 'H:\DOCUMENTS\SAS PAPERS\MODERN PORTFOLIO SAS OR';

/****************************************************************************

****************************************************************************/

%MACRO MPT_OPT(FILE,BUDGET,RETURN_VALUE,TYPE);

/*EXTRACTION OF CONTENTS*/

PROC CONTENTS DATA = &FILE(DROP = DATE) OUT = CONTENTS(KEEP = NAME) NOPRINT;
RUN;

/*SETTING THE VARIABLES*/

DATA _NULL_;
SET CONTENTS;
CALL SYMPUT('VAR'||TRIM(LEFT(_N_)),NAME);
CALL SYMPUT('VARCOUNT',TRIM(LEFT(_N_)));
RUN;

/*CALCULATING THE VARIANCE COVARIANCE TABLE*/

PROC CORR DATA = &FILE OUT=CORRTABLE(WHERE=(UPCASE(_TYPE_) IN ("COV","MEAN"))) COV NOSIMPLE
NOPRINT;
VAR
	%DO I = 1 %TO &VARCOUNT;
		&&VAR&I
	%END;
;
WITH
	%DO I = 1 %TO &VARCOUNT;
		&&VAR&I
	%END;
;
RUN;

/*COVARIANCE TABLE AND MEAN TABLE*/

DATA COVTABLE MEANTABLE;
SET CORRTABLE;
IF UPCASE(_TYPE_) IN ("MEAN") THEN OUTPUT MEANTABLE;
ELSE OUTPUT COVTABLE;
RUN;

/*ASSIGNING THE COVARIANCE VALUES*/

DATA _NULL_;
SET COVTABLE(DROP = _TYPE_ _NAME_);

ARRAY X{&VARCOUNT}
	%DO I = 1 %TO &VARCOUNT;
		&&VAR&I
	%END;
;

DO I = 1 TO &VARCOUNT;

	CALL SYMPUT('COEFF'||TRIM(LEFT(_N_))||TRIM(LEFT(I)),X{I});

END;

RUN;

/*ASSIGNING THE MEAN VALUES*/
```

Modern Portfolio Theory using SAS OR,continued

```
DATA _NULL_;
SET MEANTABLE(DROP = _TYPE_ _NAME_);

ARRAY X{&VARCOUNT}
        %DO I = 1 %TO &VARCOUNT;
                &&VAR&I
        %END;
;

DO I = 1 TO &VARCOUNT;

        CALL SYMPUT('MEAN'||TRIM(LEFT(I)),X{I});

END;

RUN;

/***********************************************************************/

/*OPTIMIZATION SECTION*/

PROC OPTMODEL;

VAR  X{1..&VARCOUNT} >= 0;

NUM COEFF{1..&VARCOUNT, 1..&VARCOUNT} = [
        %DO I = 1 %TO &VARCOUNT;
                %DO J = 1 %TO &VARCOUNT;
                        &&COEFF&I&J
                %END;
        %END;
];

NUM R{1..&VARCOUNT}=[
                %DO I = 1 %TO &VARCOUNT;
                        &&MEAN&I
                %END;
];

/* MINIMIZE THE VARIANCE OF THE PORTFOLIO'S TOTAL RETURN */
MINIMIZE F = SUM{I IN 1..&VARCOUNT, J IN 1..&VARCOUNT}COEFF[I,J]*X[I]*X[J];

/* SUBJECT TO THE FOLLOWING CONSTRAINTS */
CON BUDGET: SUM{I IN 1..&VARCOUNT}X[I] = &BUDGET;
CON GROWTH: SUM{I IN 1..&VARCOUNT}R[I]*X[I] >= &RETURN_VALUE;

%IF &TYPE ^= S %THEN %DO;
        SOLVE WITH QP;

        PRINT X;
%END;
%ELSE %DO;
        FOR {I IN 1..&VARCOUNT} X[I].LB=-X[I].UB;

        SOLVE WITH QP;
%END;

PRINT X;

QUIT;
/**/

%MEND;
```